\documentclass[13pt]{cernprep}
\usepackage{graphicx}

\begin{document}
\begin{titlepage}
\date{20 November 2015} 
\vspace{1cm}
\begin{center}
\title{\Large{The Gamma Factory  proposal for CERN$\,^\dagger$}}
\end{center}

\vspace*{2cm}

\begin{abstract} 
This year, 2015, marks the centenary of the publication of EinsteinÕs Theory of General
Relativity and it  has been named the International Year of Light  and light-based technologies by the UN General Assembly 
It is thus timely to discuss  the possibility  of broadening the present CERN research program by  including 
a new component  based on a  novel  concept of the  light source which could pave a way towards a multipurpose Gamma Factory. 
The proposed  light source could  be realized at CERN by using  the  infrastructure of the existing accelerators. 
It could  push the   intensity limits of the presently operating  light-sources by at least 7 orders of magnitude, 
reaching the flux of  the order of $10^{17}$ photons/s,  in the particularly interesting $\gamma$-ray  energy domain of \mbox{ $1 \le E_{photon} \le 400$  MeV.}   
This domain is out of  reach for the FEL-based light sources. The energy-tuned,  quasi-monochromatic gamma  beams,  
together with the gamma-beams-driven secondary beams of polarized positrons, polarized muons, neutrons and radioactive ions
would constitute the basic research tools of the proposed Gamma Factory. 
The Gamma Factory could  open new research opportunities at CERN in a vast domain of uncharted 
fundamental physics and industrial application territories. It  could 
strengthen  the leading role of CERN in the high energy frontier research territory by providing  
the unprecedented-brilliance secondary beams of polarized muons for the TeV-energy-scale muon collider  
and for the polarized-muon-beam based neutrino factory. 

\end{abstract}
\vfill  \normalsize
\begin{center}
Mieczyslaw Witold Krasny$^{\star}$
\vspace*{3mm} 

LPNHE, Universit\'{e}s Paris VI et VII and CNRS--IN2P3, Paris, France \\

\end{center}
\vspace*{5mm}
\rule{0.9\textwidth}{0.2mm}

\begin{flushleft}
\begin{footnotesize}
\hspace{15mm}
$^\dagger$ An Executive Summary of the proposal addressed to the CERN management.\\
\hspace{15mm} 
$^\star$e-mail: krasny@lpnhe.in2p3.fr
\end{footnotesize}
\end{flushleft}

\end{titlepage}



\section{Existing and future MeV-range light sources}

The light sources in the discussed MeV energy  range  have already been constructed and operating in 
several countries: HI$\gamma$S-USA, LEPS-Japan, LADON-Italy, ROKK-1-Russia, GRAAL-France and LEGS-USA. 
The  leading future project entering the construction phase is the European Union project  ELI-NP.
The ELI-NP facility is expected to produce  the flux of $10^{13}$ photons/s with the maximal energy of 20 MeV. 
The highest photon flux which has been achieved so far  is  $10^{10}$ photons/s.

All the above facilities generate, or are expected to generate,  the photon beams by the process of the 
inverse Compton scattering of the laser photons on the highly relativistic electron beams.
Since the cross section of the inverse Compton process  is small, in the \mbox{$\cal O$(1 barn)} range,
in order to achieve  the quoted above  fluxes,  the  laser system and the energy recovery linac 
technologies have to be pushed to their technological limits.  

\section{The Gamma Factory proposal  for CERN -- the leap into the ultra-high gamma-source intensity}  

The idea underlying the Gamma Factory  proposal is to use Partially Stripped Ion (PSI) beams,  
instead of  electron beams,  as the drivers of its  light source\footnote{For the discussion of the light sources 
based on PSI beams see e.g. \cite{Bessonov} and the references quoted therein}.
The PSI beams are the beams of ions carrying  
one, or more,  electrons which have not been  stripped along  the way from the ion source to the final PSI 
beam storage ring.  In the process of the resonant  absorption of the laser photons by the PSI beam,   
followed by a spontaneous  atomic-transition emissions of secondary photons, 
the initial laser-photon frequency is boosted by a factor of up to $4 \times \gamma _L ^2$, where $\gamma _L$ 
is the Lorenz factor of the partially stripped ion beam. Therefore,  the light  source in the energy range 
of  \mbox{$1 \le E_{photon} \le 400$ MeV}   
must  be driven by the high-$\gamma _L$,  LHC-stored,  PSI beams. CERN is a unique place 
in the world where such a light source could be realized. 

The cross-section for the resonant absorption of laser photons by the atomic systems is in the {\bf giga-barn range}, 
while the cross-section for the point-like electrons is in  the {\bf barn range}. As a consequence the  PSI-beam-driven light source
intensity could be higher than those of the electron-beam-driven ones by a large factor. 
For the light source working in the  regime of multiple  photon emissions by each of the beam ion,  
the  photon beam intensity is expected to be  limited no longer by the laser light intensity but 
by the available RF power of the ring in which 
partially stripped ions are stored.  For example, the flux of up to $10^{17}$ photons/s
could  be achieved for photon energies in the 10 MeV region already with the present, U= 16 MV,  circumferential voltage of the LHC cavities. 
This photon flux is by a  factor of $10^7$  higher  than that of the highest-intensity  electron-beam-driven light source, HI$\gamma$S@Durham,  
operating in the  same  energy regime. 

If photon beams carrying more than \mbox{$\cal O$(100 kW)} of beam-power can be safely handled,  and  
if the present circumferential voltage could be  increased  (at LEP2 the corresponding value was 3560 MV),
even higher fluxes could  be  generated.

\section{Acceleration, storage and use of the PSI  beams at CERN}

The first steps to understand the storage stability of the PSI beams were already made at BNL. The $^{77+}$Au  
beam with two unstripped electrons was successfully  circulating  in the AGS ring at BNL and, more recently, in its  RHIC ring \cite{Dejan}. 
These tests may be considered as  a starting  point for further exploratory tests which could be  
carried out initially at the CERN SPS and, if successful, at the LHC. 

If stable PSI beams could be produced  and stored  they 
would not only drive the photon source but could also be used for the following two unconventional applications. 

Firstly, they  would allow the LHC to operate  as an {\bf electron-proton(ion) collider} \cite{Krasny}.  
The LHC experiments could simply record collisions of electrons, brought to LHC experiment's  interaction  points 
"on the shoulders" of the  ion-carriers,   with the counter-propagating proton(ion) beam.  

Secondly, they may turn out to be efficient  driver beams  for the hadron beam driven 
plasma-wakefield acceleration \cite{Caldwell} of a witness beam. This is  because the  PSI bunches, contrary to the proton bunches,  
could be very efficiently cooled by the Doppler laser cooling techniques,  allowing to compress their bunch sizes. 
A profit could thus be made from the fact that the maximal achievable plasma 
electric field acceleration gradient  increases  quadratically with the decreasing bunch length of the driver beam. 

It remains to be stressed that  a large fraction of the beam cooling and beam manipulation techniques 
exploiting  the  internal degrees of freedom of the beam particles, 
which have been mastered over the three decades by the atomic physics community, 
could be directly applied to  the  high energy PSI beams.

\section{The photon collision schemes and secondary beams of the Gamma Factory } 

The high intensity and high brilliance gamma beam could be used to realize, for the first time,  a {\bf photon-photon collider at CERN}:
(1) in the range of CM energies of 1 - 100 KeV, for  collisions  of the gamma beam with the laser photons,  
and (2) in the energy range of 1 -  800 MeV,  for the gamma beam collisions with the  counter propagating,  twin gamma beam.  

The gamma beam could also collide with the LHC proton and fully stripped ion beams, The CM energy range of the corresponding  {\bf photon-proton} 
and {\bf photon-nucleus colliders} \ would be  4 - 60 GeV.  

Finally, the gamma beam  could  be extracted from the LHC  and used to produce high intensity  secondary beams of:

\begin{itemize} 
\item  
{\bf Polarized electrons and positrons with the expected intensity which could exceed $\mathbf{10^{17}}$ positrons/s}. 
Such an intensity would be three orders of magnitude  higher than that of the KEK positron source and largely 
satisfy the source requirements for both the ILC and CLIC colliders, and even that of  a future high luminosity  ep (eA) collider project based 
on the energy recovery linac. 

\item 
{\bf Polarized muons with the expected intensity of $\mathbf{10^{12}}$ muons/s,  
and the tertiary neutrino beams of the corresponding intensity}.
The above muon  intensity would be  four orders of magnitude higher than that of  the Paul Scherrer Institute's "$\pi$E5" muon beam. It would  
satisfy the intensity requirements for the $L = 10^{34}$ s$^{-1}$cm$^{-2}$, 3 TeV muon collider.  
The corresponding neutrino fluxes would be a factor of 50 higher than those of nuSTORM. Moreover,  thanks to the initial muon polarization 
the muon-neutrino (muon-anti-neutrino) beams  could be uncontaminated by the electron-neutrino (electron-anti-neutrino) contributions.
The neutrino and antineutrino bunches could be separated with 100 \% efficiency on the bases of their timing. 
In addition, the fluxes could be predicted to a very high accuracy, providing 
an optimal  neutrino-beam configuration  for the high systematic precision measurements e.g. of  the CP-violating phase in the neutrino CKM
matrix. To reach even higher  muon (neutrino) intensities two paths  could be envisaged. In the first one,  
based on the conversion of the  high energy gamma beam into muon pairs, the present circumferential voltage of 
the LHC  would have to be upgraded and a specialized design of the gamma conversion targets would have to be made. 
An alternative scheme would be to tune  the gamma beam energy to a significantly lower energy -- just above the electron-positron 
pair production threshold, reducing thus both the circumferential voltage and the beam power strains. 
The positron bunches, produced by such a low energy gamma beam, 
would need to be  accelerated in the dedicated positron ring to the energy exceeding the muon pair production threshold 
in collisions with the stationary target electrons, $E_e \sim  (2 m_{\mu}^2)/(m_e)$.  The intensity of the muon beam  produced 
in such a scheme could be increased by replacing the single-pass collisions of the positron beam by the multipass collisions\cite{Raimondi}.
For both the above two types of muon beams  the product of the beam longitudinal and transverse 
emittances could  be at least four orders of magnitude  smaller than that for the pion-decay-originated muon source.  
The basic obstacle in forming  the small-emittance muon-beam for the future muon colliders 
and the future long baseline neutrino program,  which is inherent to the 
pion-decay generated muon beams -- the necessity of the intensive initial muon beam cooling --  
could  be   circumvented for  the Gamma Factory driven muon beams.

\item 
{\bf Neutrons   with the  expected intensity reaching $\mathbf{10^{15}}$ neutrons/s (first generation  neutrons) and radioactive, neutron-rich  ions with the 
intensity reaching $\mathbf{10^{14}}$ ions/s}. Preliminary estimates show that the intensity of the Gamma Factory beams of neutrons and radioactive ions  
could approach those of the planned
future European projects  such as the ESS and EURISOL. The Gamma Factory beams  may turn out  to be  more effective in terms of 
their power consumption efficiency since almost 10 \% of the LHC RF power could be  converted into the power of  the neutron and radioactive ion beams
if  the energy of the photon beam  is tuned to  the Giant Dipole Resonance (GDR) region of the target nuclei.

\end{itemize} 

New Gamma Factory  beam lines of  unprecedented intensities  and its high luminosity photon-photon, photon-proton 
and photon-nucleus collision interaction points could  attract new scientific communities to CERN. This  could lead 
not only to a diversification of the CERN-based  scientific program but also, if necessary,  to  an enlargement  of the CERN funding resources  
which could be crated by the use of the specialized  beam lines for the industrial and medical applications.   

\section{Expected highlights of the Gamma Factory research program } 

The  physics research domains which could be explored  by this proposal include:
fundamental QED measurements (for example, for the first time, the  elastic light-light scattering 
could  be observed with the rate of $\approx$1000 events/s);
dark matter searches (mainly via the dark photon and neutron portals); 
investigation of basic symmetries of the  Universe (neutron dipole moment, neutron-antineutron
oscillations, forbidden muon decays);
studies of color  confinement;
nuclear photonics; 
physics of neutron-rich radioactive beams,  
physics with energy-tagged  neutron beam
and the vast domain of the atomic physics of muonic and electronic atoms. 

As far as the CERN flagship high energy frontier research is concerned, 
the Gamma Factory high brilliance beams of polarized positron and muons could pave the way 
to execute at CERN the research programs of: (1)  {\bf a  TeV-energy-scale muon collider};  
(2) {\bf neutrino factory}, and (3)  {\bf a TeV-energy-scale muon-hadron collider}. 

The CERN Gamma Factory project could open a wide spectrum of industrial and medical applications
in the following domains:
muon catalyzed cold fusion;
gamma-beam catalyzed hot fusion;
Accelerator Driven System (ADS)  and Energy Amplifier (EA)  research;
nondestructive assay and segregation of nuclear waste;
transmutation of nuclear waste;
material studies of thick objects and 
production of ions for Positron Emission Tomography (PET)  and 
for the selective cancer-cell therapy with alpha emitters. 

\section{The way forward}

The presented above research option  may turn out not only to be scientifically attractive but also cost-effective
because it proposes to re-use,  in a novel manner, the existing CERN accelerator infrastructure. It  
may be considered as complementary to the present  hadron-collision program
and could be performed at any stage of the LHC life-time.  
 
In oder to prove that such a future option is not only a conceptually attractive but also a viable one two 
initial exploratory  paths may be considered and initiated early on. 

The goal of the first one would be  to perform a detailed  validation  of the achievable performance figures 
of the Gamma Factory initiative  for each  branch  of its application domains, 
to build up  the physics case for its research  program and,  most importantly,  to attract a wide community to this initiative.

The goal of the second one would be  to prove experimentally the concepts underlying this proposal. Most of 
the feasibility tests can be performed at the SPS and organized such that  the ongoing CERN research program is hardly affected.   
The necessary experimental tests could start with  special SPS runs with partially stripped ions to optimize ion-stripping schemes, to
measure the PSI beam life-time and  beam emittance evolution. The next step would include 
a short test run with single PSI bunches injected to the LHC to measure  the beam life-time, beam losses and  beam emittances.
To complete the feasibility proof,  these tests would need to be followed  by a "proof-of-principle"  SPS experiment. The  initial idea,  
which remains to be validated by the outcome of the SPS PSI beam stability tests,  would be 
to collide  the SPS Nitrogen-like Xenon beam with  the Krypton 645 nm laser photons and to 
measure the properties of the generated X-ray beam. 
Such an experiment could be done, in its  initial phase 
in the  North Hall using the extracted PSI beam beam and subsequently,  in the  
specially designed (e.g. in the UA2 cavern) laser-PSI-beam collision point in the SPS tunnel. 

To enter  these  two exploratory paths and to elaborate them  initial discussions
with CERN experts are extremely important. Their primary goal would be to define all the elements of 
the requisite studies, their timing and milestones.

\section{The support for such an initiative} 

The proposal outlined  in this short note can be realized  only at CERN. 
It  will not take off  as a project without an  endorsement and support of 
the CERN scientific community  and of the CERN management. 
The  principal goal of this note is to expose  the Gamma Factory  proposal
for the initial consideration and to ask for an institutional support for this initiative already at its present conceptual stage.
A start-up support could involve setting up of a working group based at CERN for exploratory studies including accelerator physicists, 
choosing  its chairperson(s) and organizing an initial workshop at CERN to gather  both 
the accelerator physicists and potential Gamma Factory  users  to discuss the scientific and technological interests and merits of the 
project.


\end{document}